\documentstyle[12pt,graphicx]{article}
\begin{document}
\baselineskip 0.6cm

\def\simgt{\mathrel{\lower2.5pt\vbox{\lineskip=0pt\baselineskip=0pt
           \hbox{$>$}\hbox{$\sim$}}}}
\def\simlt{\mathrel{\lower2.5pt\vbox{\lineskip=0pt\baselineskip=0pt
           \hbox{$<$}\hbox{$\sim$}}}}

\begin{titlepage}

\begin{flushright}
SLAC-PUB-11487 \\
UCB-PTH-05/29 \\
LBNL-58876
\end{flushright}

\vskip 1.7cm

\begin{center}

{\Large \bf 
Dark Matter before the LHC in a Natural Supersymmetric Standard Model
}

\vskip 1.0cm

{\large
Ryuichiro Kitano$^a$ and Yasunori Nomura$^{b,c}$}

\vskip 0.4cm

$^a$ {\it Stanford Linear Accelerator Center, 
                Stanford University, Stanford, CA 94309} \\
$^b$ {\it Department of Physics, University of California,
                Berkeley, CA 94720} \\
$^c$ {\it Theoretical Physics Group, Lawrence Berkeley National Laboratory,
                Berkeley, CA 94720} \\

\vskip 1.2cm

\abstract{We show that the solid lower bound of about $10^{-44}~{\rm cm}^2$ 
is obtained for the cross section between the supersymmetric dark matter 
and nucleon in a theory in which the supersymmetric fine-tuning problem is 
solved without extending the Higgs sector at the weak scale.  This bound 
arises because of relatively small superparticle masses and a fortunate 
correlation that the two dominant diagrams for the dark matter detection 
always interfere constructively if the constraint from the $b \rightarrow 
s \gamma$ measurements is obeyed.  It is, therefore, quite promising in 
the present scenario that the supersymmetric dark matter is discovered 
before the LHC, assuming that the dark matter is the lightest 
supersymmetric particle.}

\end{center}
\end{titlepage}

Weak scale supersymmetry provides an elegant framework to address the 
naturalness problem of the standard model as well as to provide the dark 
matter of the universe.  On the other hand, there are already strong 
constraints on how supersymmetry can be realized at the weak scale.  One 
of the severest constraints comes from the LEP~II bound on the Higgs boson 
mass~\cite{Barate:2003sz}.  To evade this bound, the top squarks should 
generically be heavy, which in turn gives a large negative correction 
to the Higgs mass-squared parameter.  This requires fine-tuning of order 
a few percent or worse to reproduce the correct scale for electroweak 
symmetry breaking in many supersymmetric models.  Since our main 
motivation for weak scale supersymmetry comes from the concept of 
naturalness, we do not want fine-tuning in a theory with supersymmetry. 
This problem is called the supersymmetric fine-tuning problem. 
Addressing this issue may provide a key to find the correct theory 
at the weak scale, but it is generically difficult to solve the problem 
without extending the structure of the Higgs sector substantially. 

Recently, we have shown that the supersymmetric fine-tuning problem 
can be solved within the framework of the minimal supersymmetric 
standard model (MSSM)~\cite{Kitano:2005wc}.  Crucial ingredients for 
the solution are a large $A$ term for the top squarks and a small 
$B$ term for the Higgs doublets, which ensure successful electroweak 
symmetry breaking with relatively small top squark masses.  A one-loop 
correction from the top-stop loop to the Higgs mass-squared parameter 
is made small by effectively lowering the messenger scale of supersymmetry 
breaking~\cite{Choi:2005uz,Choi:2005hd}, by invoking a mixture of 
moduli~\cite{Kaplunovsky:1993rd} and anomaly mediated~\cite{Randall:1998uk} 
contributions as a source of supersymmetry breaking~\cite{Choi:2004sx}. 
An interesting aspect of this solution is that all the superparticles 
are relatively light, with the masses determined only by a few free 
parameters.  To evade fine-tuning worse than $20\%$, we find that the 
gauginos, whose masses are almost universal at the weak scale, must be 
lighter than about $900~{\rm GeV}$.  The squarks and sleptons are even 
lighter, with the masses a factor of $\sqrt{2}$ smaller than that of 
the gauginos.  Such light superparticles can have large impacts on the 
detections of these particles at colliders and in other experiments. 

In this paper we show that the detection of supersymmetric dark matter 
is quite promising in the framework described above for solving the 
supersymmetric fine-tuning problem.  In this framework, the lighter 
neutral Higgsino is a good candidate for the dark matter.  First, it 
is lighter than about $200~{\rm GeV}$ from the naturalness requirement, 
and thus is the lightest supersymmetric particle (LSP).  Second, it 
can be generated nonthermally by the decay of the gravitino, which 
in turn is generated by the decay of the moduli field.  For the 
range of the gravitino and moduli masses relevant to us, the right 
abundance of $\Omega_{\tilde{h}^0} \simeq 0.2$ can be naturally 
obtained~\cite{Kohri:2005ru}.  A comparable amount of relic Higgsinos 
may also arise from the decay of Q-balls formed in the early stage 
of baryogenesis~\cite{Fujii:2002kr}.  In any event, once we have the 
Higgsino as the dominant component of the dark matter, we can calculate 
its detection rates independently of the production mechanism. 

We find that in the parameter region relevant to our solution to the 
fine-tuning problem, the cross sections between the Higgsino and nuclei 
are dominated by the $t$-channel Higgs boson exchange diagrams.  There 
are two contributions coming from the lighter and heavier Higgs boson 
exchanges, which are generically comparable in size.  We find that the 
two contributions interfere constructively (destructively) if the sign 
of the $\mu$ parameter is positive (negative) in the standard convention. 
On the other hand, there is a strong constraint on the sign of $\mu$ 
from the measurements of the $b \rightarrow s \gamma$ process.  We find 
that if the sign of $\mu$ is positive a large portion of the parameter 
space is consistent with the current measurements, while the case with 
negative $\mu$ is almost completely excluded.  Because of this fortunate 
correlation, together with the fact that Higgsino dark matter generically 
has larger cross sections to nuclei than bino dark matter, the detection 
of supersymmetric dark matter in our framework is quite promising. 
We find that the parameter region relevant for the solution to the 
fine-tuning problem gives the dark matter-nucleon spin-independent 
cross section ranging from about $10^{-44}~{\rm cm}^2$ to several 
times $10^{-42}~{\rm cm}^2$.  Some of the parameter region, therefore, 
is even already excluded by the latest data from the CDMS~II 
experiment~\cite{Akerib:2005kh}.  Since we obtain the lower bound 
on the cross section of about $10^{-44}~{\rm cm}^2$, we expect that 
most of the relevant parameter space will be covered by further 
running of the CDMS~II in the next two years.

Let us now start calculating the cross section between the Higgsino 
dark matter and nuclei in the present framework.  For this purpose, we 
need the masses and interactions of the superparticles and the Higgs 
bosons.  The soft supersymmetry breaking parameters for the gauginos 
and sfermions are given by
\begin{equation}
  M_1 = M_2 = M_3 = M_0,
\label{eq:gaugino-masses}
\end{equation}
\begin{equation}
  m_{\tilde{q}}^2 = m_{\tilde{u}}^2 = m_{\tilde{d}}^2 
    = m_{\tilde{l}}^2 = m_{\tilde{e}}^2 = \frac{M_0^2}{2},
\label{eq:sfermion-masses}
\end{equation}
\begin{equation}
  A_u = A_d = A_e = M_0,
\label{eq:Aterms}
\end{equation}
at the effective messenger scale $M_{\rm mess} \sim {\rm TeV}$, where 
the parameter $M_0$ takes a value in the range
\begin{equation}
  450~{\rm GeV} \simlt M_0 \simlt 900~{\rm GeV},
\label{eq:M0-range}
\end{equation}
for $\tan\beta \equiv \langle H_u \rangle/\langle H_d \rangle \simgt 20$. 
For smaller $\tan\beta$, the lower bound in Eq.~(\ref{eq:M0-range}) 
increases; for $\tan\beta = 10$ ($5$) the lower bound becomes $\approx 
550~{\rm GeV}$ ($900~{\rm GeV}$).  The range of $\tan\beta$ we consider 
is thus
\begin{equation}
  \tan\beta \simgt 5.
\label{eq:tan-beta}
\end{equation}
The upper bound in Eq.~(\ref{eq:M0-range}) comes from requiring 
the absence of fine-tuning worse than $20\%$: $\Delta^{-1} \geq 
20\%$. The Higgs sector of the theory contains four free parameters, 
$m_{H_u}^2$, $m_{H_d}^2$, $\mu$ and $B$, but one combination is fixed 
by the vacuum expectation value $v \equiv (\langle H_u \rangle^2 + 
\langle H_d \rangle^2)^{1/2} \simeq 174~{\rm GeV}$, leaving only three 
independent parameters.  We take these parameters to be $\mu$, $m_A$ 
and $\tan\beta$, where $m_A$ is the mass of the pseudo-scalar Higgs 
boson.  The naturalness requirement, $\Delta^{-1} \geq 20\%$, then 
leads to the bounds
\begin{equation}
  |\mu| \simlt 190~{\rm GeV},
\label{eq:mu-range}
\end{equation}
and
\begin{equation}
  m_A \simlt 300~{\rm GeV}.
\label{eq:mA-range}
\end{equation}
Overall, the number of free parameters in our analysis is five: $M_0$, 
$M_{\rm mess}$, $\tan\beta$, $\mu$ and $m_A$.  The values of $M_0$, 
$\tan\beta$, $\mu$ and $m_A$ are bounded by Eqs.~(\ref{eq:M0-range}), 
(\ref{eq:tan-beta}), (\ref{eq:mu-range}) and (\ref{eq:mA-range}), 
respectively, and $M_{\rm mess}$ has to be close to TeV: $50~{\rm GeV} 
\simlt M_{\rm mess} \simlt 3~{\rm TeV}$. 

From Eqs.~(\ref{eq:gaugino-masses}~--~\ref{eq:M0-range}) and 
(\ref{eq:mu-range}), we find that the Higgsinos, which have masses of 
about $|\mu|$, are significantly lighter than the other superparticles, 
so that the LSP comes from one of the Higgsinos.  After electroweak 
symmetry breaking, there are three Higgsino states: two neutral states 
$\tilde{h}^0$ and $\tilde{h}'^0$, defined here by $m_{\tilde{h}'^0} 
> m_{\tilde{h}^0}$, and one charged state $\tilde{h}^{\pm}$.  The mass 
splittings between these states are approximately given by
\begin{equation}
  m_{\tilde{h}'^0} - m_{\tilde{h}^{\pm}} 
    \simeq m_{\tilde{h}^{\pm}} - m_{\tilde{h}^0}
    \simeq \frac{m_Z^2}{2 M_0},
\label{eq:Higgsinos}
\end{equation}
where the corrections are only of order $20\%$ in the relevant parameter 
region.  Therefore, we find that the lighter neutral Higgsino is always 
the LSP, and the mass splittings between the different Higgsino states 
are of order $10~{\rm GeV}$. 

Suppose now that the neutral Higgsino $\tilde{h}^0$, produced 
nonthermally in the early universe, is the dominant component of 
the dark matter.  Since the mass splittings in Eq.~(\ref{eq:Higgsinos}) 
are much larger than the kinetic energy of the dark matter, we only 
have to consider elastic scattering between $\tilde{h}^0$ and nuclei 
to study the detection of the Higgsino dark matter.  The scattering 
occurs through exchanges of the squarks and Higgs bosons, but in the 
parameter region relevant to us the Higgs boson exchange diagrams 
dominate.  We thus focus on the Higgs boson exchange contributions 
below.  The squark exchange contributions, however, are also included 
in our numerical analysis given later.

The exchange of the Higgs bosons induces the following effective 
interactions between the dark matter $\chi$ and the quarks 
$q = u,d,s,c,b,t$~\cite{Drees:1993bu}:
\begin{equation}
  {\cal L}_{\rm eff} = \sum_{q} f_q m_q \bar{\chi} \chi \bar{q} q,
\label{eq:eff-L}
\end{equation}
where $m_q$ are the quark masses and $f_q$ are given by
\begin{equation}
  f_q \simeq - \frac{g^2+g'^2}{8(M_0-|\mu|)} 
    \Biggl( \frac{c_{h\chi\chi}c_{hqq}}{m_h^2} 
      + \frac{c_{H\chi\chi}c_{Hqq}}{m_H^2} \Biggr).
\label{eq:fq}
\end{equation}
Here, $m_h$ and $m_H$ are the masses of the lighter and heavier neutral 
Higgs bosons, $h$ and $H$, and the coefficients $c$'s are given by
\begin{equation}
  c_{h\chi\chi} \simeq 1 + {\rm sgn}(\mu)\sin 2\beta,
\qquad
  c_{H\chi\chi} \simeq - {\rm sgn}(\mu) \cos 2\beta,
\label{eq:c-hcc}
\end{equation}
\begin{equation}
  c_{hqq} = \left\{ \begin{array}{ll} 
    \frac{\cos\alpha}{\sin\beta}  \; & ({\rm for}\;\; q=u,c,t) \\
    -\frac{\sin\alpha}{\cos\beta} \; & ({\rm for}\;\; q=d,s,b)
  \end{array} \right.,
\qquad
  c_{Hqq} = \left\{ \begin{array}{ll} 
    \frac{\sin\alpha}{\sin\beta} \; & ({\rm for}\;\; q=u,c,t) \\
    \frac{\cos\alpha}{\cos\beta} \; & ({\rm for}\;\; q=d,s,b)
  \end{array} \right.,
\label{eq:c-hqq}
\end{equation}
where $\alpha$ is the mixing angle for the neutral Higgs bosons 
($-\pi/2 \leq \alpha \leq 0$).  The spin-independent cross section 
of the dark matter with a target nucleus is then given by
\begin{equation}
  \sigma_T = 
    \frac{4}{\pi} \Biggl( \frac{m_\chi m_T}{m_\chi+m_T} \Biggr)^2 
    (n_p f_p + n_n f_n)^2,
\label{eq:chi-nuclei}
\end{equation}
where $m_T$ is the mass of the target nucleus, and $n_p$ and $n_n$ 
are the proton and neutron numbers in the nucleus, respectively. 
The quantities $f_p$ and $f_n$ are given by
\begin{equation}
  f_N = m_N \sum_q f_q {\cal M}^{(N)}_q,
\label{eq:FN}
\end{equation}
where $N=p,n$.  Here, ${\cal M}^{(N)}_q$ are the nucleon matrix 
elements defined by $\langle N |\, m_q \bar{q} q\, | N \rangle 
= m_N {\cal M}^{(N)}_q$, whose approximate values are given by
\begin{eqnarray}
  {\cal M}^{(p)}_u \simeq 0.023, \quad
  {\cal M}^{(p)}_d \simeq 0.034, \quad
  {\cal M}^{(p)}_s \simeq 0.14, \quad
  {\cal M}^{(p)}_c \simeq {\cal M}^{(p)}_b 
    \simeq {\cal M}^{(p)}_t \simeq 0.059,
\label{eq:Ms-u} \\
  {\cal M}^{(n)}_u \simeq 0.019, \quad
  {\cal M}^{(n)}_d \simeq 0.041, \quad
  {\cal M}^{(n)}_s \simeq 0.14, \quad
  {\cal M}^{(n)}_c \simeq {\cal M}^{(n)}_b 
    \simeq {\cal M}^{(n)}_t \simeq 0.059.
\label{eq:Ms-d}
\end{eqnarray}

\ From these expressions, we find the following.  First, 
Eq.~(\ref{eq:c-hcc}) tells us that the signs of $c_{h\chi\chi}$ 
and $c_{H\chi\chi}$ are given by ${\rm sgn}(c_{h\chi\chi}) = +1$ 
and ${\rm sgn}(c_{H\chi\chi}) = {\rm sgn}(\mu)$, respectively. 
On the other hand, from Eqs.~(\ref{eq:FN}~--~\ref{eq:Ms-d}), we find 
that the total contributions to the cross section from operators 
with down-type quarks, $q=d,s,b$, are larger than those from operators 
with up-type quarks, $q=u,c,t$.  Combining these with the fact that 
${\rm sgn}(c_{hqq}) = {\rm sgn}(c_{Hqq}) = +1$ for $q=d,s,b$ (see 
Eq.~(\ref{eq:c-hqq})), we find from Eq.~(\ref{eq:fq}) that the 
contributions from the light and heavy Higgs bosons interfere 
constructively (destructively) if the sign of $\mu$ is positive 
(negative).  This implies that for $\mu > 0$ we have a solid 
lower bound on the cross section because our mass parameters $M_0$, 
$\mu$, $m_h$, and $m_H$ are bounded by Eq.~(\ref{eq:M0-range}), 
Eq.~(\ref{eq:mu-range}), $m_h \simlt 120~{\rm GeV}$, and $m_H \simeq 
m_A \simlt 300~{\rm GeV}$ (see Eq.~(\ref{eq:mA-range})), respectively. 
Here, the bound on $m_h$ follows from that on $M_0$.  The case with 
$\mu < 0$, on the other hand, admits a possibility of cancellation 
between the two contributions.  Since $\alpha \sim \beta-\pi/2$ in 
our parameter region, an excessive cancellation occurs if $m_H^2/m_h^2 
\sim \tan\beta$ (see Eqs.~(\ref{eq:fq},~\ref{eq:c-hcc},~\ref{eq:c-hqq})). 

\begin{figure}[t]
\begin{center}
  \includegraphics[height=7.5cm]{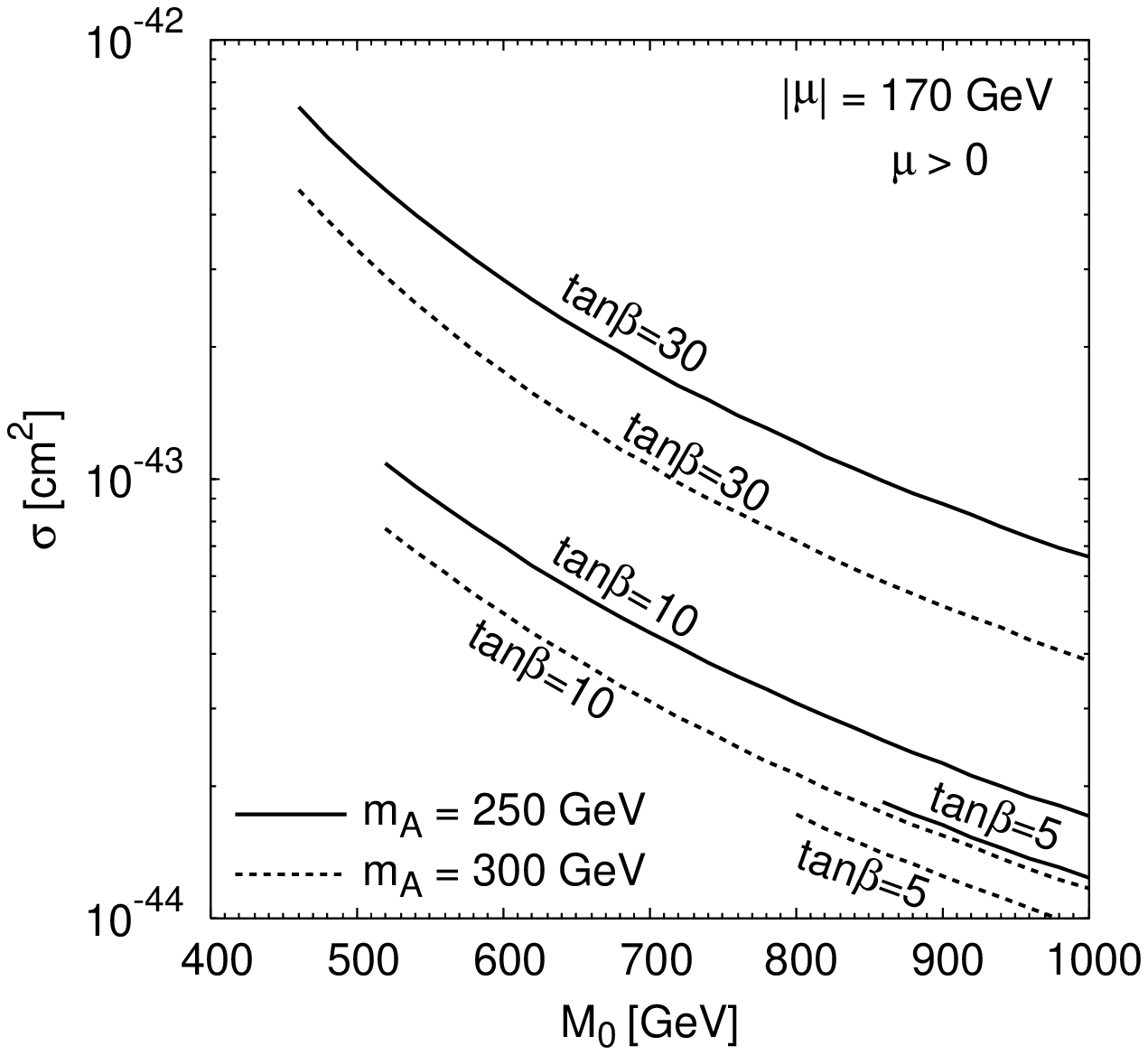}
  \includegraphics[height=7.5cm]{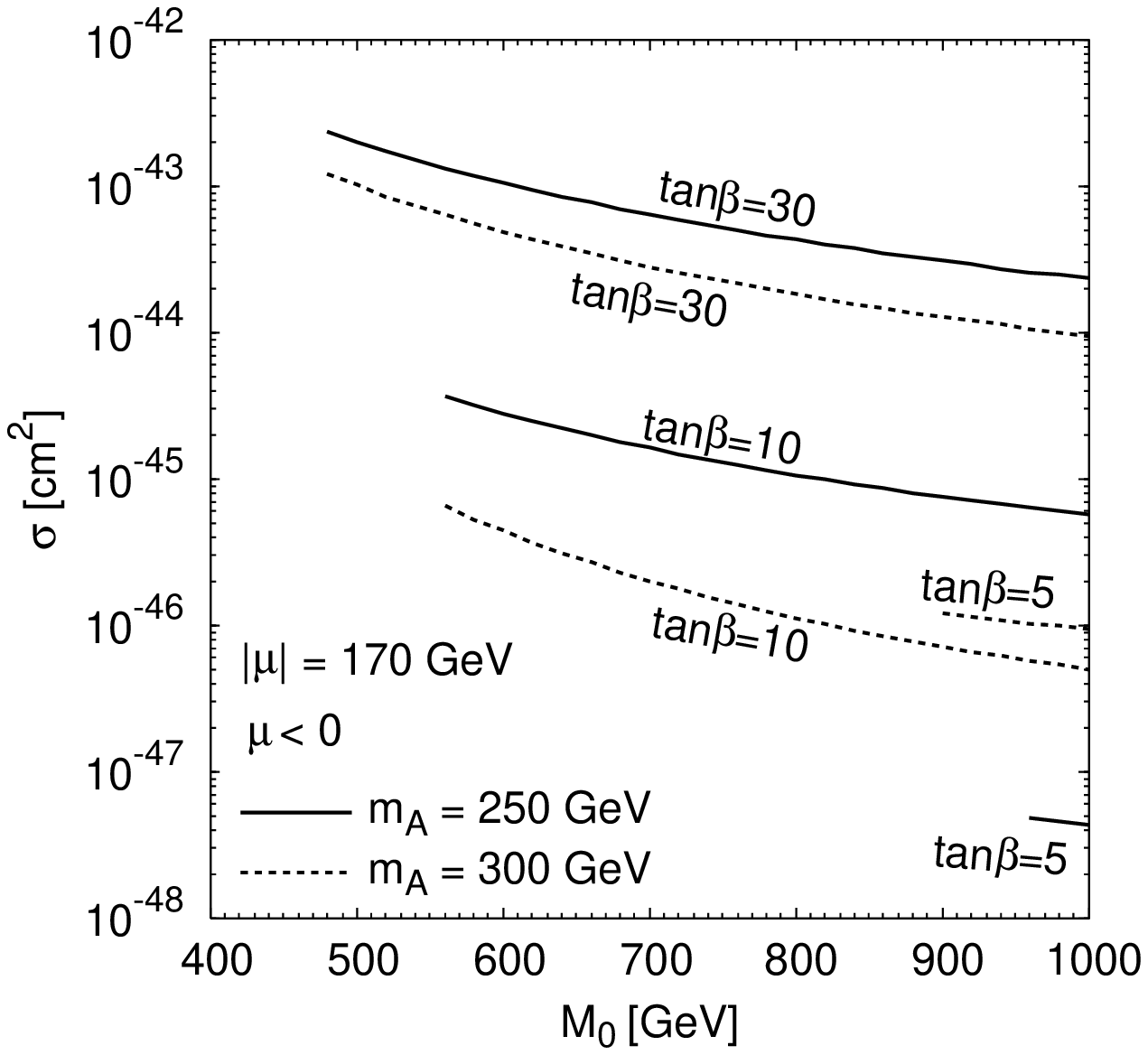}
\end{center}
\caption{Dark matter-nucleon cross section, $\sigma$, as a function 
 of $M_0$ for the case of $X = {\rm Ge}$.  The left and right plots 
 correspond to the cases with $\mu > 0$ and $\mu < 0$, respectively. 
 The parameters $|\mu|$ and $M_{\rm mess}$ are fixed as $|\mu| 
 = 170~{\rm GeV}$ and $M_{\rm mess} = 1~{\rm TeV}$, and $\tan\beta$ 
 and $m_A$ are varied as $\tan\beta = 5, 10, 30$ and $m_A = \{ 250, 
 300 \}~{\rm GeV}$.  The curves are drawn only for the values of $M_0$ 
 for which the Higgs boson mass bound of $m_h \simgt 114.4~{\rm GeV}$ 
 is satisfied.}
\label{fig:cross-section}
\end{figure}

In Fig.~\ref{fig:cross-section}, we have plotted the cross section 
of the dark matter $\chi$ with a target nucleus $X$ in the case of 
$X = {\rm Ge}$, for $\mu > 0$ (left) and $\mu < 0$ (right). 
Here, we have plotted the cross section normalized to the 
nucleon~\cite{Lewin:1995rx}
\begin{equation}
  \sigma = \frac{\sigma_T}{(n_p+n_n)^2}
    \Biggl( \frac{1~{\rm GeV}}{m_\chi m_T/(m_\chi+m_T)} \Biggr)^2,
\label{eq:sigma}
\end{equation}
as a function of $M_0$ for several values of $\tan\beta$ and $m_A$: 
$\tan\beta = 5, 10, 30$ and $m_A = \{ 250, 300 \}~{\rm GeV}$.  In the 
figure, all the contributions discussed in~\cite{Drees:1993bu} including 
the squark exchange contributions, as well as the effects of running 
from $M_{\rm mess}$ to the superparticle mass scale, are included.  The 
curves are drawn only for the values of $M_0$ for which the LEP~II bound 
on the Higgs boson mass, $m_h \simgt 114.4~{\rm GeV}$~\cite{Barate:2003sz}, 
is satisfied.  The parameters $|\mu|$ and $M_{\rm mess}$ are fixed as 
$|\mu| = 170~{\rm GeV}$ and $M_{\rm mess} = 1~{\rm TeV}$, but the 
sensitivities to these parameters are not significant. 

In the figure, we can clearly see the general features discussed before. 
First, the cross section is larger for $\mu > 0$ than $\mu < 0$.  Second, 
we obtain the lower bound on $\sigma$ of about $10^{-44}~{\rm cm}^2$ for 
$\mu > 0$, since $M_0$ is bounded as $M_0 \simlt 900~{\rm GeV}$ from the 
naturalness requirement.  This is quite interesting because this region 
will be probed in the further running of the CDMS~II experiment in the 
next two years.  On the other hand, we find that the cross section can 
be as small as $(10^{-47}\!\sim\!10^{-46})~{\rm cm}^2$ for $\mu < 0$. 
Therefore, fully exploring this case would require next generation 
experiments. 

\begin{figure}[t]
\begin{center}
  \includegraphics[height=7.5cm]{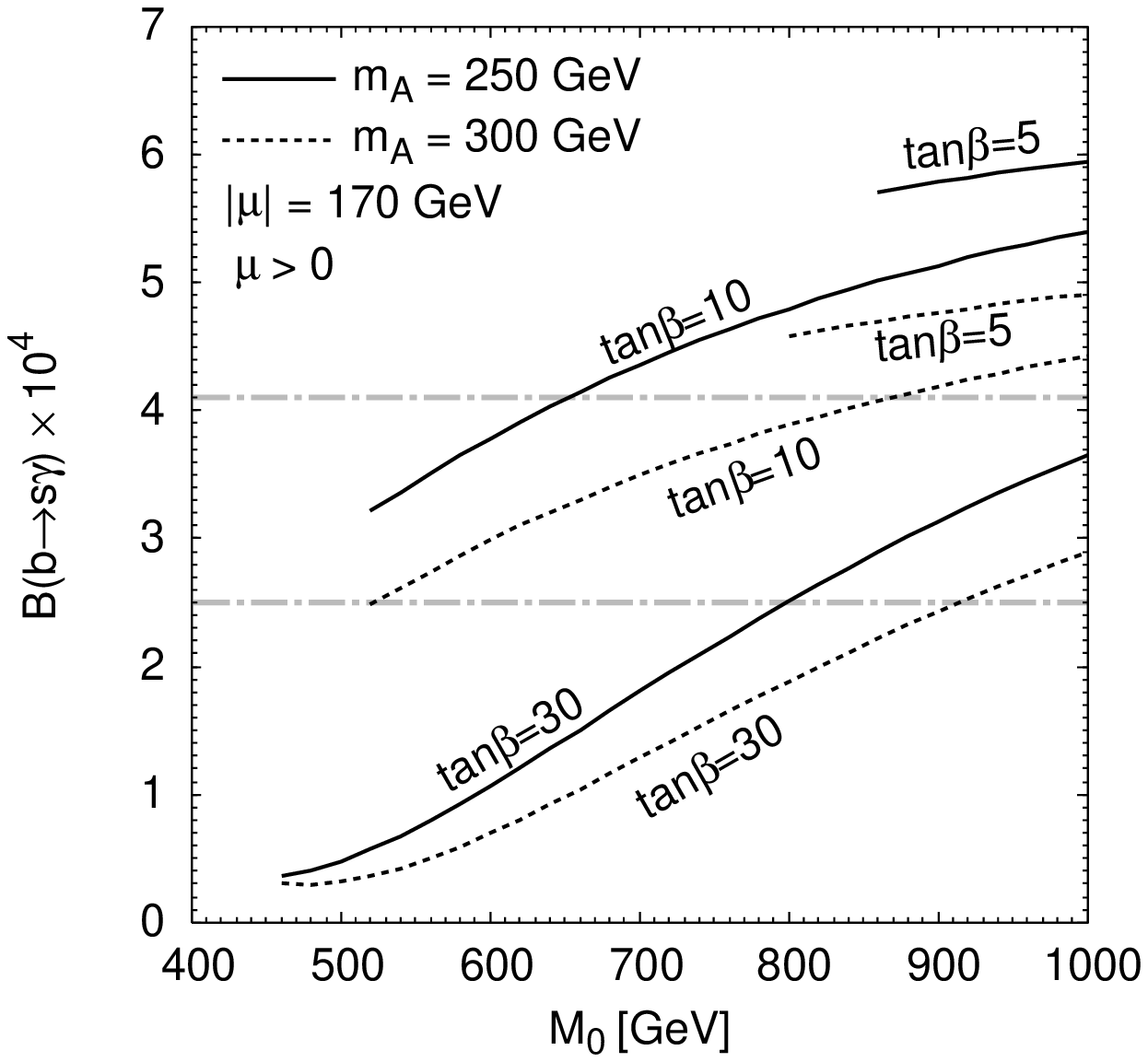}
  \includegraphics[height=7.5cm]{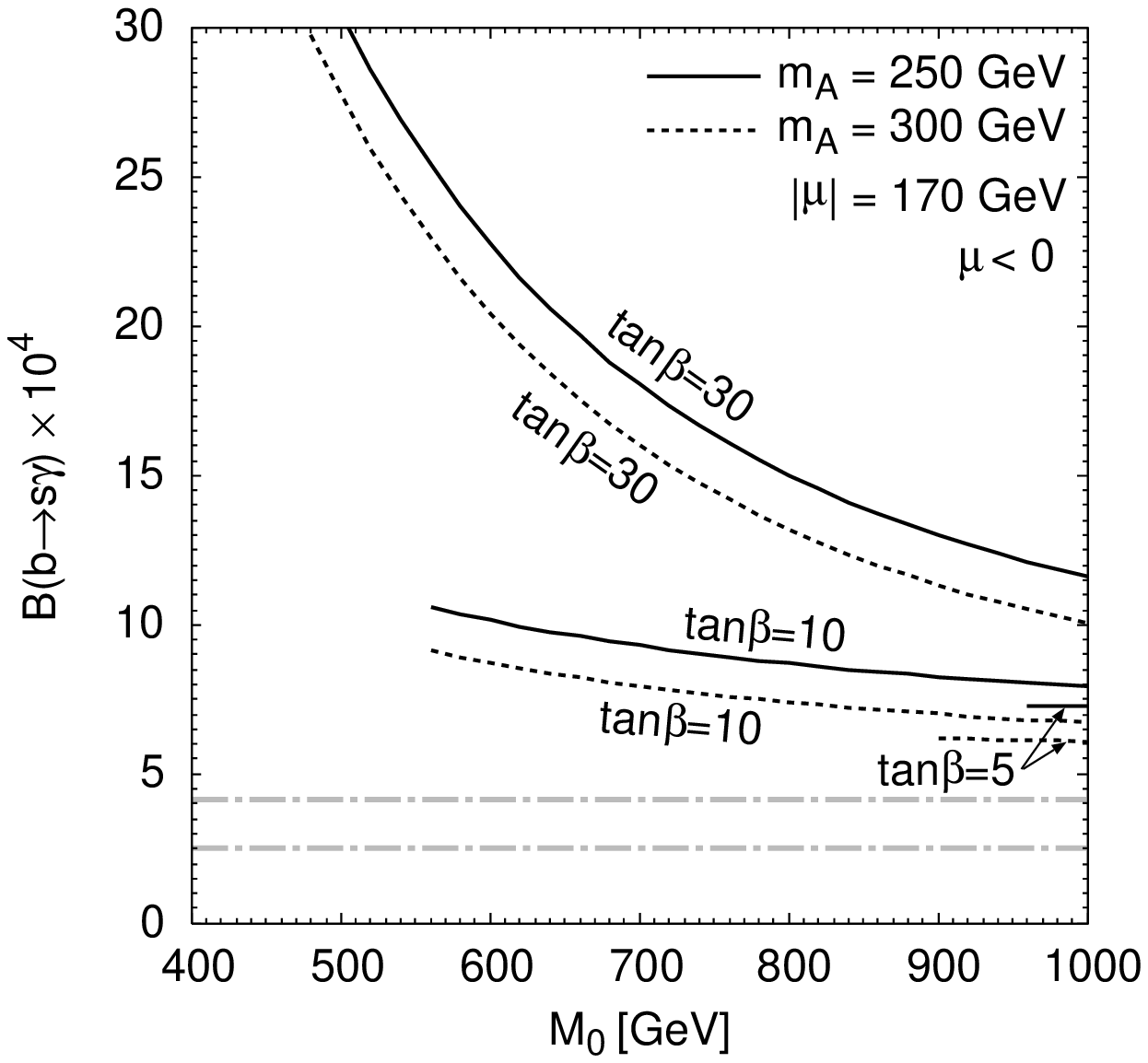}
\end{center}
\caption{Branching ratio for the $b \rightarrow s \gamma$ process, 
 ${\rm B}(b \rightarrow s \gamma)$, as a function of $M_0$.  The left 
 and right plots correspond to the cases with $\mu > 0$ and $\mu < 0$, 
 respectively.  The parameters $|\mu|$ and $M_{\rm mess}$ are fixed 
 as $|\mu| = 170~{\rm GeV}$ and $M_{\rm mess} = 1~{\rm TeV}$, and 
 $\tan\beta$ and $m_A$ are varied as $\tan\beta = 5, 10, 30$ and 
 $m_A = \{ 250, 300 \}~{\rm GeV}$.  The curves are drawn only for 
 the values of $M_0$ for which the Higgs boson mass bound of $m_h 
 \simgt 114.4~{\rm GeV}$ is satisfied.}
\label{fig:b-s-gamma}
\end{figure}

Can we say something about the sign of $\mu$ from other experiments? 
It is known that the rate for the $b \rightarrow s \gamma$ process can 
depend highly on the signs of $\mu$ and $A_t$.  In our parameter region, 
contributions from chargino and charged Higgs boson loops interfere 
destructively (constructively) for $\mu > 0$ ($< 0$) in the 
$b \rightarrow s \gamma$ amplitude, implying that the positive sign 
of $\mu$ is preferred over the other one.  This situation is depicted 
in Fig.~\ref{fig:b-s-gamma}, where we have plotted the branching ratio 
${\rm B}(b \rightarrow s \gamma)$ for $\mu > 0$ (left) and $\mu < 0$ 
(right), calculated using the DarkSUSY package~\cite{Gondolo:2004sc}.
We have again drawn the curves as a function of $M_0$ for 
$\tan\beta = 5,10,30$, $m_A = \{ 250, 300 \}~{\rm GeV}$, $|\mu| 
= 170~{\rm GeV}$ and $M_{\rm mess} = 1~{\rm TeV}$ only for the 
regions where the bound on the Higgs boson mass is satisfied.  The 
sensitivities to the fixed parameters, $\mu$ and $M_{\rm mess}$, are 
weak in our parameter region.  We have also indicated the $2\sigma$ 
allowed region from the current measurements, ${\rm B}(b \rightarrow 
s \gamma) = (3.3 \pm 0.4) \times 10^{-4}$~\cite{Eidelman:2004wy}, 
by two horizontal dot-dashed lines.

While the predictions for the rate of the $b \rightarrow s \gamma$ 
process are still subject to theoretical uncertainties of order 
$(20\!\sim\!30)\%$~\cite{Gambino:2001ew}, we can still conclude 
from Fig.~\ref{fig:b-s-gamma} that the case with $\mu < 0$ is almost 
entirely excluded.  On the other hand, for $\mu > 0$ a large portion 
of the parameter space is consistent with the data within theoretical 
and experimental uncertainties.  The positive sign of $\mu$ may also 
be preferred from the measurement of the muon anomalous magnetic 
moment~\cite{Bennett:2004pv}, because supersymmetric corrections 
decrease (increase) the possible discrepancy between the standard 
model prediction and the data for $\mu > 0$ ($< 0$).

\begin{figure}[t]
\begin{center}
  \includegraphics[height=7.5cm]{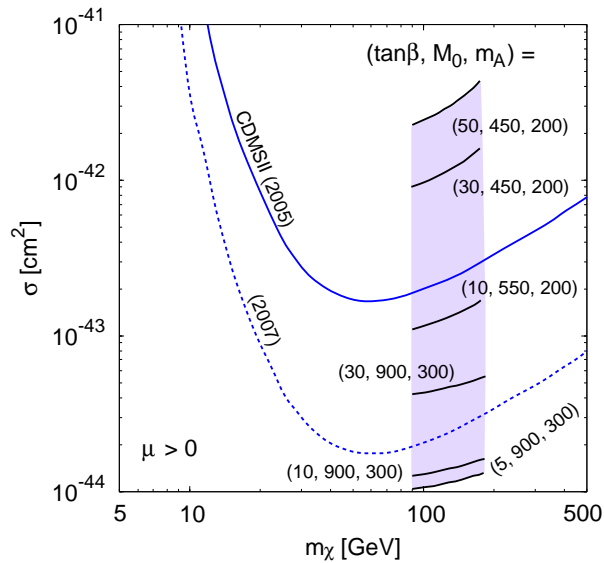}
\end{center}
\caption{Dark matter-nucleon cross section, $\sigma$, as 
 a function of the dark matter mass $m_{\chi}$ for various values 
 of $(\tan\beta, M_0, m_A)$.  The shaded area corresponds to our 
 parameter region in which the supersymmetric fine-tuning problem 
 is solved.  The exclusion curve from the latest CDMS~II data is also 
 shown (a solid line), as well as an expected future sensitivity 
 (a dashed line).}
\label{fig:CDMS}
\end{figure}

As we have seen before, the detection of our Higgsino dark matter is 
quite promising for $\mu > 0$.  In Fig.~\ref{fig:CDMS}, we have plotted 
the spin-independent cross section between the dark matter and nucleon 
as a function of the dark matter mass $m_\chi$ for various values of 
$\tan\beta$, $M_0$ and $m_A$.
For $\tan\beta = 10$ ($30$), the values of $M_0$ and $m_A$ are chosen 
such that they maximize, $(M_0, m_A) = (550, 200)$ ($(M_0, m_A) = 
(450, 200)$), or minimize, $(M_0, m_A) = (900, 300)$ ($(M_0, m_A) = 
(900, 300)$), the cross section, so that we can read off the allowed 
range of the cross section for a given value of $\tan\beta$ and $m_\chi$. 
(The lower bound on $m_A$ has been set to $200~{\rm GeV}$ somewhat 
arbitrarily, motivated by the $b \rightarrow s \gamma$ data, but lowering 
this only extends the range of $\sigma$ to the direction of larger 
$\sigma$.)  The lines of $(\tan\beta, M_0, m_A) = (50, 450, 200)$ and 
$(5, 900, 300)$ have been drawn to set the upper and lower boundaries 
for the range of $\sigma$.  Together with the experimental lower bound 
on $m_\chi$ and the naturalness bound on $\mu$ in Eq.~(\ref{eq:mu-range}), 
we obtain our predictions for $\sigma$ and $m_\chi$ given by the shaded 
area in the figure.  We have also drawn the exclusion curve from the 
latest CDMS~II data~\cite{Akerib:2005kh} by a solid line, and an 
estimate for the expected future sensitivity (an order of magnitude 
improvement of the current bound by the end of 2007~\cite{CDMS}) by 
a dashed line.

It is interesting that the latest data from the CDMS~II already 
exclude a portion of parameter space relevant for our solution to the 
supersymmetric fine-tuning problem.  For $\tan\beta = 30$, for example, 
the region $M_0 \simlt 750~{\rm GeV}$ ($500~{\rm GeV}$) is already 
excluded for $m_A = 200~{\rm GeV}$ ($300~{\rm GeV}$).  The current data 
are not yet very sensitive to smaller values for $\tan\beta$.  A large 
portion of the relevant parameter space, however, will be covered by the 
end of 2007 by a planned order of magnitude improvement of the current 
lower bound on $\sigma$.  For $\tan\beta = 30$, for example, this 
covers the entire parameter region relevant for the solution to the 
supersymmetric fine-tuning problem.  Even for $\tan\beta = 10$, the 
entire region of $M_0$ is covered for $m_A = 200~{\rm GeV}$, and 
$M_0 \simlt 700~{\rm GeV}$ is covered for $m_A = 300~{\rm GeV}$. 
It is, therefore, quite promising in the present framework that the 
supersymmetric dark matter is discovered before the LHC, assuming 
that the dominant component of the dark matter is in fact the 
lightest supersymmetric particle.

In summary, we have shown that the detection of supersymmetric dark 
matter is quite promising in the theory in which the supersymmetric 
fine-tuning problem is solved.  In fact, this conclusion is somewhat 
more generic in the sense that in a theory where the supersymmetric 
fine-tuning is reduced, the Higgsino is generically light and is 
naturally a candidate for the LSP.  Then, if the gaugino masses are 
not very large, as suggested indirectly by the fine-tuning argument, 
the detection rate is generically not very small.  To obtain the 
solid lower bound on the detection rate, however, we need more 
information.  We have shown that the $b \rightarrow s \gamma$ process 
can provide such information in our particular context, and that we 
can obtain the lower bound on the dark matter-nucleon cross section, 
$\sigma \simgt 10^{-44}~{\rm cm}^2$, which is close to the expected 
sensitivity of the CDMS~II experiment within the next two years. 
We find it quite interesting that the supersymmetric dark matter 
may in fact be discovered before the LHC turns on.

\section*{Acknowledgments}

Y.N. thanks Jeffrey Filippini for conversation.  The work of R.K. 
was supported by the U.S. Department of Energy under contract number 
DE-AC02-76SF00515.  The work of Y.N. was supported in part by the 
Director, Office of Science, Office of High Energy and Nuclear Physics, 
of the US Department of Energy under Contract DE-AC02-05CH11231, by 
the National Science Foundation under grant PHY-0403380, by a DOE 
Outstanding Junior Investigator award, and by an Alfred P. Sloan 
Research Fellowship.

\end{document}